\documentclass[aps,prd,onecolumn,groupedaddress,showpacs,nofootinbib,amssymb]{revtex4}
\usepackage[dvips]{graphicx}
\usepackage{amssymb}
\usepackage{amsmath}
\usepackage{graphicx,,color}
\usepackage{amsfonts}
\usepackage{bm}

\newcommand\be{\begin{equation}}
\newcommand\ee{\end{equation}}

\allowdisplaybreaks[4]

\begin{document}

\title{Cosmology in a model with Lagrange multiplier, and Gauss-Bonnet and \\
 non-minimal kinetic couplings}
\author{D. F. Jimenez,$^{1,2}$\,\thanks{mail1}
L. N. Granda,$^{2}$\,\thanks{mail2} E. Elizalde,$^{1}$\thanks{mail1}}
\affiliation{$^{1)}$  Institute of Space Sciences (IEEC-CSIC) C. Can Magrans s/n,
08193 Bellaterra, Barcelona, Spain\\
$^{2)}$ Departamento de f\'isica, Universidad del Valle, A.A. 25360 Cali, Colombia\\
}

\tolerance=5000

\begin{abstract}
A scalar-tensor model with Gauss-Bonnet and non-minimal kinetic couplings is considered, in which ghost modes are eliminated via a Lagrange multiplier constraint. A reconstruction procedure is deviced for the scalar potential and  Lagrange multiplier, valid for any given cosmological scenario. In particular,  inflationary and  dark energy cosmologies of different types (power-law, Little-Rip, de Sitter, quasi de Sitter) are reconstructed in such models.  It is shown that, for various choices of the kinetic coupling terms, it is possible to obtain a viable inflationary phenomenology compatible with the most accurate values of the observational indices.
\end{abstract}

\maketitle

\section{Introduction}

The observation, contrary to all theoretical models, of a clear diminution in the observed energy fluxes coming from type Ia supernovae \cite{perlmutter,riess,kowalski,hicken}, as compared  with the expected values, opened one of the most puzzling and severe problems in cosmology today. These observations have been now interpreted as solid evidence for an accelerating phase in the universe, whose energy balance is being dominated by something called dark energy. The accelerated expansion is presently supported by astrophysical data from several independent sources, as the cosmic microwave background anisotropy \cite{wmap}, large scale galaxy surveys, and baryon acoustic oscillations of the matter density power spectrum \cite{percival}, among others. In addition to the late-time acceleration, there are compelling reasons to believe that the universe experienced another phase of sudden acceleration at verey early times, called inflation \cite{Guth:1980zm,Starobinsky:1982ee,Linde:1983gd}, which simultaneously solves the horizon, flatness, homogeneity and density of monopoles problems, which had troubled the standard Big Bang model of cosmology. The standard inflationary paradigm is realized by using a homogeneus scalar field, dubbed inflaton, with a potential dominating the energy density of the universe (for reviews, see \cite{inflation1,inflation2,inflation4}). Although the standard inflation is produced by a minimally coupled scalar field, it is possible to obtain a valid inflationary phase within other models, such as kinetic inflation \cite{picon}, non-minimal derivative couplings \cite{cgermani,lgranda}, and string theory inspired inflation \cite{kallosh}. 

A quite successfull line of work that has sought to resolve the inflation and dark energy problems are modified gravity  theories. In this respect, \cite{reviews1} presents a general review of the latest developments in the description of the inflationary era, dark energy, unified representation of inflation and late-time acceleration, particularly in the context of modified $f(R)$, $f({\cal{G}})$, and $f(T)$ gravity theories (for other reviews on modified gravity, see e.g. \cite{reviews2,reviews3}). In \cite{FRG}  a unified cosmology was proposed, with early and late-time dynamics controlled by $F(R,\cal{G})$ gravity, thus able to produce both inflation and the dark energy era, the intermediate era being approximately identical to the standard Einstein-Hilbert gravity. A realistic unification of early-time inflation with late-time dark energy was first proposed  in \cite{nuevaser1}, and was developed in detail, in reasonable versions of $F(R)$ gravity, in \cite{nuevaser2,nuevaser3,nuevaser4,nuevaser5}. A model with a coupling between the scalar field and the Gauss-Bonnet (GB) $\cal{G}$ invariant, to study inflationary slow-roll cosmology,  was proposed in \cite{viableinflation}, displaying a phenomenology viable for a wide range of values of the free parameters of the theory, and  novel bottom-up reconstruction techniques from the observational indices. Many studies of inflation with this type of coupling have appeared in the literature (for an important stream of related papers see, for example, \cite{Yi:2018gse,Guo:2009uk,Guo:2010jr,Jiang:2013gza,Koh:2014bka,Koh:2016abf,Kanti:2015pda,vandeBruck:2017voa,Kanti:1998jd,Nozari:2017rta,Chakraborty:2018scm}). This same coupling  has also been proposed to address the dark energy problem in \cite{Nojiri:2005vv}, where it was found that quintessence or a phantom phase may occur in the late time universe. Different aspects of these accelerating cosmologies with GB corrections have been discussed in \cite{tsujikawan1,leith,koiviston1,koiviston2,neupanen1,Nojiri:2006je,Calcagni:2005im,Cognola:2006sp}, and a modified GB theory applied to dark energy was suggested in \cite{nojirin1}.  Late time cosmological solutions in a model that includes both nonminimal kinetic coupling to curvature and a coupling of the scalar field to the GB invariant, have been studied in \cite{granda1,granda2,granda3}. The problem of the ghost degrees of freedom in GB modified gravity theories has been studied in \cite{ghostfree}, where it was demonstrated that the Lagrange multiplier formalism leads to the elimination of the ghost modes in both  $f({\cal{G}})$  and $F(R,\cal{G})$ gravity theories.

In this paper we will examine an extension of the ghost-free Gauss-Bonnet model \cite{ghostfree} by adding an explicit non-minimal kinetic coupling to curvature. In Section II, we review the main equations in a general background, in the flat FRW metric, and give a detailed reconstruction scheme for a given cosmological scenario.  In Section III we calculate the main observational indexes corresponding to inflation, namely the spectral index of the primordial curvature perturbations and the tensor-to-scalar ratio, and we obtain a viable phenomenology, compatible with the most recent observational data.

\section{Action with Lagrange multiplier and Gauss-Bonnet and non-minimal kinetic couplings}

\subsection{Field Equations}

In this section we shall briefly review the formalism with a Lagrange multiplier. We adhere to the notation and presentation of Ref. \cite{ghostfree}. Consider the following ghost-free action with field Lagrange multiplier $\lambda$, and with Gauss-Bonnet and non-minimal kinetic couplings, namely
\begin{equation}\label{actiongeneral}
S = \int {{d^4}x\sqrt { - g} \left[ {\frac{R}{{2{\kappa ^2}}} + \lambda \left( {\frac{1}{2}{\partial _\mu }\chi {\partial ^\mu }\chi  + \frac{{{\mu ^4}}}{2}} \right) - \frac{1}{2}{\partial _\mu }\chi {\partial ^\mu }\chi  - V(\chi ) + h(\chi ){\cal{G}} + {F_1}(\chi ){G_{\mu \nu }}{\partial ^\mu }\chi {\partial ^\nu }\chi } \right]},
\end{equation}
where ${G_{\mu \nu }} = {R_{\mu \nu }} - \frac{1}{2}{g_{\mu \nu }}R$ is the Einstein tensor and the new function $F_{1}(\chi)$ has dimension of ${(length)^2}$. The coupling $h(\chi)$ is dimensionless and the constant $\mu$ has mass-dimension one. By varying the above action  with respect to $\lambda$,  we obtain  the constraint
\begin{equation}\label{restriccion}
0 = \frac{1}{2}{\partial _\mu }\chi {\partial ^\mu }\chi  + \frac{{{\mu ^4}}}{2},
\end{equation}
which allows to redefine the scalar potential $V (\chi)$ as follows,
\[\tilde V(\chi ) \equiv \frac{1}{2}{\partial _\mu }\chi {\partial ^\mu }\chi  + V(\chi ) =  - \frac{{{\mu ^4}}}{2} + V(\chi ).\] 
This allows us to rewrite the action of Eq. (\ref{actiongeneral}) as 
\begin{equation}\label{modelser}
S = \int {{d^4}x\sqrt { - g} \left[ {\frac{R}{{2{\kappa ^2}}} + \lambda \left( {\frac{1}{2}{\partial _\mu }\chi {\partial ^\mu }\chi  + \frac{{{\mu ^4}}}{2}} \right) - \tilde V(\chi ) + h(\chi ){\cal{G}} + {F_1}(\chi ){G_{\mu \nu }}{\partial ^\mu }\chi {\partial ^\nu }\chi } \right]}. 
\end{equation}
By varying Eq. (\ref{modelser}) with respect to the metric and the scalar field, we derive the gravitational field equations given by the expressions
\begin{equation}\label{eqcamposer}
 - \frac{1}{{{\kappa ^2}}}{G_{\mu \nu }} - \lambda {\partial _\mu }\chi {\partial _\nu }\chi  - {g_{\mu \nu }}\tilde V(\chi ) + T_{\mu \nu }^{GB} + T_{\mu \nu }^K = 0,
\end{equation}
and
\begin{equation}\label{eqmovser}
 - {\nabla _\mu }\left[ {\lambda {\nabla ^\mu }\chi } \right] - \tilde V'(\chi ) + h'(\chi )G - {F_1}'(\chi ){G_{\mu \nu }}{\partial ^\mu }\chi {\partial ^\nu }\chi  - 2{F_1}(\chi ){G_{\mu \nu }}{\nabla ^\mu }{\nabla ^\nu }\chi  = 0,
\end{equation}
where $ T_{\mu \nu }^K$ corresponds to the variation of the kinetic coupling, and $ T_{\mu \nu }^{GB}$ comes from the variation of the coupling to GB. These variations can be written, respectively, as
\begin{align}\label{GB}
T_{\mu \nu }^{GB}&= 4\Big([{\nabla _\mu }{\nabla _\nu }{h}(\chi )]R - {g_{\mu \nu }}[{\nabla _\rho }{\nabla ^\rho }{h}(\chi )]R - 2[{\nabla ^\rho }{\nabla _\mu }{h}(\chi )]{R_{\nu \rho }} - 2[{\nabla ^\rho }{\nabla _\nu }{h}(\chi )]{R_{\mu \rho }} \nonumber \\
\nonumber \\[-0.5cm]
 &{\kern 1pt}  + 2[{\nabla _\rho }{\nabla ^\rho }{h}(\chi )]{R_{\mu \nu }} + 2{g_{\mu \nu }}[{\nabla ^\rho }{\nabla ^\sigma }{h}(\chi )]{R_{\rho \sigma }} - 2[{\nabla ^\rho }{\nabla ^\sigma }{h}(\chi )]{R_{\mu \rho \nu \sigma }}\Big),
\end{align}
and
\begin{align}\label{K}
T_{\mu \nu }^K &=G_{\mu \nu }{F_1}(\chi ){\nabla _\lambda }\chi {\nabla ^\lambda }\chi  + {g_{\mu \nu }}{\nabla _\lambda }{\nabla ^\lambda }\left( {{F_1}(\chi ){\nabla _\gamma }\chi {\nabla ^\gamma }\chi } \right) -{{\nabla _\nu }{\nabla _\mu }}\left( {{F_1}\left( \chi  \right){\nabla _\lambda }\chi {\nabla ^\lambda }\chi } \right)\nonumber \\
\nonumber \\[-0.5cm]
& + R{F_1}\left( \chi  \right){\nabla _\mu }\chi {\nabla _\nu }\chi  - 2{F_1}\left( \chi  \right)\left( {{R_{\mu \lambda }}{\nabla ^\lambda }\chi {\nabla _\nu }\chi  + {R_{\nu \lambda }}{\nabla ^\lambda }\chi {\nabla _\mu }\chi } \right)+ {g_{\mu \nu }}{R_{\lambda \gamma }}{F_1}\left( \chi  \right){\nabla ^\lambda }\chi {\nabla ^\gamma }\chi \nonumber \\
\nonumber \\[-0.5cm]
&+{\nabla _\lambda }{\nabla _\mu }\left( {{F_1}\left( \chi  \right){\nabla ^\lambda }\chi {\nabla _\nu }\chi } \right) + {\nabla _\lambda }{\nabla _\nu }\left( {{F_1}\left( \chi  \right){\nabla ^\lambda }\chi {\nabla _\mu }\chi } \right)- {\nabla _\lambda }{\nabla ^\lambda }\left( {{F_1}\left( \chi  \right){\nabla _\mu }\chi {\nabla _\nu }\chi } \right)\nonumber \\
\nonumber \\[-0.5cm]
&  - {g_{\mu \nu }}{\nabla _\lambda }{\nabla _\gamma }\left( {{F_1}\left( \chi  \right){\nabla ^\lambda }\chi {\nabla ^\gamma }\chi } \right).
\end{align}

\subsection{FRW cosmology and reconstruction}
Let us consider the spatially-flat Friedmann-Lemaitre-Robertson-Walker (FLRW) metric
\begin{equation}
\label{metricfrw} ds^2 = - dt^2 + a(t)^2 \sum_{i=1,2,3}
\left(dx^i\right)^2\,.
\end{equation}
 We also assume that $\lambda$ and $\chi$ depend solely on the cosmic time $t$, that is, $\lambda=\lambda(t)$ and $\chi=\chi(t)$. Then, a solution of Eq. (\ref{restriccion}) is given below
\begin{equation}\label{restriccionfrw}
\chi  = {\mu ^2}t.
\end{equation}
The (00) and (11) components of the Eq. (\ref{eqcamposer}) take the form (with the Hubble parameter being $H=\dot{a}/a$)
\begin{equation}\label{1friser}
 - \frac{{3{H^2}}}{{{\kappa ^2}}} + \tilde V(\chi ) + 9{F_1}(\chi ){H^2}{\mu ^4} - \lambda {\mu ^4} - 24{H^3}{\mu ^2}h'(\chi ) = 0
\end{equation}
and
\begin{equation}\label{2friser}
\frac{{3{H^2}}}{{{\kappa ^2}}} + \frac{{2\dot H}}{{{\kappa ^2}}} - \tilde V(\chi ) - {\mu ^4}{F_1}(\chi )(3{H^2} + 2\dot H) - 2H{\mu ^6}{F_1}'(\chi ) + 16{H^3}{\mu ^2}h'(\chi ) + 16H\dot H{\mu ^2}h'(\chi )+ 8{H^2}{\mu ^4}h''(\chi ) = 0. 
\end{equation}
On the other hand, Eq. (\ref{eqmovser}) yields
\begin{equation}\label{contser}
{\mu ^2}\dot \lambda  + 3H\lambda {\mu ^2} - {\mu ^2}H{F_1}(\chi )(18{H^2} + 12\dot H) - 3{H^2}{\mu ^4}{F_1}'(\chi ) + 24{h^\prime }(\chi ){H^2}({H^2} + \dot H) - \tilde V'(\chi ) = 0.
\end{equation}
Eqs. (\ref{1friser}) and (\ref{2friser}) can be solved with respect to $\tilde V(\chi )$ and $\lambda$, as follows,
\begin{equation}\label{Vser}
\tilde V(\chi ) = \frac{{3{H^2}}}{{{\kappa ^2}}} + \frac{{2\dot H}}{{{\kappa ^2}}} - 3{F_1}(\chi ){H^2}{\mu ^4} - 2H{\mu ^6}{F_1}'(\chi ) + 16{H^3}{\mu ^2}h'(\chi ) - 2{F_1}(\chi )\dot H{\mu ^4} + 16H\dot H{\mu ^2}h'(\chi ) + 8{H^2}{\mu ^4}h''(\chi )
\end{equation}
and
\begin{equation}\label{lser}
\lambda  = 6{F_1}(\chi ){H^2} - 2{\mu ^2}H{F_1}'(\chi ) - \frac{{8{H^3}h'(\chi )}}{{{\mu ^2}}} + \frac{{2\dot H}}{{{\kappa ^2}{\mu ^4}}} - 2{F_1}(\chi )\dot H + \frac{{16H\dot Hh'(\chi )}}{{{\mu ^2}}} + 8{H^2}h''(\chi ),
\end{equation}
where in both Eqs. (\ref{Vser}) and (\ref{lser}), the functions $F_{1}(\chi)$ and $h(\chi)$  are arbitrary. As a consequence, any cosmological scenario encoded in the Hubble parameter, $H(t)$, can be realized for this model. It is worth exemplifying the reconstruction technique here; so let us demonstrate how this method works, by choosing some explicit cases. 

As first example, we consider a de Sitter spacetime realization, in which case the Hubble rate $H$ is constant, $H = H_0$. Then, by using Eqs. (\ref{Vser}) and (\ref{lser}), for arbitrarily chosen functions $F_{1}(\chi)$ and $h(\chi)$, the corresponding scalar potential and Lagrange multiplier $\lambda$ are given by,
\[\begin{gathered}
\tilde  V(\chi ) = \frac{{3{H_0}^2}}{{{\kappa^2}}} - 3{H_0}^2{\mu ^4}{F_1}(\chi ) - 2{H_0}{\mu ^6}{F_1}'(\chi ) + 16{H_0}^3{\mu ^2}h'(\chi ) + 8{H_0}^2{\mu ^4}h''(\chi ), \hfill \\
  \lambda  = 6{H_0}^2{F_1}(\chi ) - 2{H_0}{\mu ^2}{F_1}'(\chi ) - \frac{{8{H_0}^3h'(\chi )}}{{{\mu ^2}}} + 8{H_0}^2h''(\chi ). \hfill \\ 
\end{gathered} \]
The next class of examples is given by the power-law solutions ($a \sim {t^p}$), which are of special interest because they represent asymptotic or intermediate states among all possible cosmological evolutions. Then, by using  Eqs. (\ref{Vser}) and (\ref{lser}), we find that 
\[\begin{gathered}
  \tilde V(\chi ) = \frac{{3{p^2} - 2p}}{{{\kappa ^2}{t^2}}} + \frac{{{\mu ^4}{F_1}(\chi )(2p - 3{p^2})}}{{{t^2}}} - \frac{{2p{\mu ^6}{F_1}'(\chi )}}{t} + \frac{{16{\mu ^2}{h^\prime }(\chi )({p^3} - {p^2})}}{{{t^3}}} + \frac{{8{p^2}{\mu ^4}h''(\chi )}}{{{t^2}}}, \hfill \\
  \lambda  =  - \frac{{2p}}{{{\kappa ^2}{t^2}{\mu ^4}}} + \frac{{2p{F_1}(\chi )}}{{{t^2}}} + \frac{{6{p^2}{F_1}(\chi )}}{{{t^2}}} - \frac{{2p{\mu ^2}{F_1}'(\chi )}}{t} - \frac{{16{p^2}{h^\prime }(\chi )}}{{{t^3}{\mu ^2}}} - \frac{{8{p^3}{h^\prime }(\chi )}}{{{t^3}{\mu ^2}}} + \frac{{8{p^2}h''(\chi )}}{{{t^2}}}, \hfill \\ 
\end{gathered} \]
where $t=\frac{{\chi}}{{\mu^2}}$. 

Another set of examples consists of the solutions known as Little Rip. This type of solutions represent an alternative to Big
Rip ones \cite{frampton,brevick}, where the dark energy density increases with time but without facing a finite time future singularity. Consider the behavior of the Hubble parameter
\[H = {H_L}{e^{\gamma t}},\]
where $H_L$ and $\gamma$ are positive constants. From the above equation, it follows that $\dot H>0$, which reproduces a phantom super-accelerated phase, free of future singularities. Then, by using  Eqs. (\ref{Vser}) and (\ref{lser}), we find
$$\tilde V(\chi ) = (3{e^{2t\gamma }}H_L^2 + 2{e^{t\gamma }}{H_L}\gamma )\left( {\frac{1}{{{\kappa ^2}}} - {\mu ^4}{F_1}(\chi )} \right) - 2{e^{t\gamma }}{H_L}{\mu ^6}{F_1}'(\chi ) + 16{\mu ^2}h'(\chi )({e^{3t\gamma }}H_L^3 + {e^{2t\gamma }}H_L^2\gamma ) + 8{e^{2t\gamma }}H_L^2{\mu ^4}h''(\chi )$$
and
$$\lambda  = \frac{{2{e^{t\gamma }}{H_L}\gamma }}{{{\kappa ^2}{\mu ^4}}} + {F_1}(\chi )(6{e^{2t\gamma }}H_L^2 - 2{e^{t\gamma }}{H_L}\gamma ) - 2{e^{t\gamma }}{H_L}{\mu ^2}{F_1}'(\chi ) - \frac{{h'(\chi )(8{e^{3t\gamma }}H_L^3 - 16{e^{2t\gamma }}H_L^2\gamma )}}{{{\mu ^2}}} + 8{e^{2t\gamma }}H_L^2h''(\chi ),$$
where $t=\frac{{\chi}}{{\mu^2}}$. 

Finally, a particularly interesting and phenomenologically valuable evolution is given by the quasi-de Sitter evolution, in which case the Hubble rate reads  \cite{reviews1} 
\begin{equation}\label{quasi}
H = {H_0} - {H_i}(t - {t_k}),
\end{equation}
where $H_0$ and $H_i$ are dimensionful parameters of the theory, and $t_k$ is the time instant when the primordial curvature modes exit the horizon. In general, these parameters are  severely constrained by the observational data. By replacing them in  Eqs. (\ref{Vser}) and (\ref{lser}), we find the following expressions for the potential and Lagrange multiplier, respectively,
\begin{align}\label{potquasi}
\tilde V(\chi )  =&  - \frac{{2{H_i}}}{{{\kappa ^2}}} + \frac{{3{{({H_0} - {H_i}(t - {t_k}))}^2}}}{{{\kappa ^2}}} + 2{H_i}{\mu ^4}{F_1}(\chi ) - 3{({H_0} - {H_i}(t - {t_k}))^2}{\mu ^4}{F_1}(\chi )- 2({H_0} - {H_i}(t - {t_k})){\mu ^6}{F_1}'(\chi ) \nonumber\\
\nonumber \\[-0.5cm]
& - 16{H_i}({H_0} - {H_i}(t - {t_k})){\mu ^2}h'(\chi ) + 16{({H_0} - {H_i}(t - {t_k}))^3}{\mu ^2}h'(\chi )+ 8{({H_0} - {H_i}(t - {t_k}))^2}{\mu ^4}h''(\chi )
\end{align}
and
\begin{align}\label{lagquasi}
\lambda  =&  - \frac{{2{H_i}}}{{{\kappa ^2}{\mu ^4}}} + 2{H_i}{F_1}(\chi ) + 6{({H_0} - {H_i}(t - {t_k}))^2}{F_1}(\chi ) - 2({H_0} - {H_i}(t - {t_k})){\mu ^2}{F_1}'(\chi ) \nonumber\\
\nonumber \\[-0.5cm]
&- \frac{{16{H_i}({H_0} - {H_i}(t - {t_k}))h'(\chi )}}{{{\mu ^2}}} - \frac{{8{{({H_0} - {H_i}(t - {t_k}))}^3}h'(\chi )}}{{{\mu ^2}}} + 8{({H_0} - {H_i}(t - {t_k}))^2}h''(\chi ).
\end{align}
In the following section we  use this scenario in the context of inflationary cosmology, and we calculate  the spectral index of the primordial curvature perturbations, $n_s$, and the scalar-to-tensor ratio, $r$.

\section{Observational inflation indexes}
\begin{figure}[h]
\centering
\includegraphics[width=21pc]{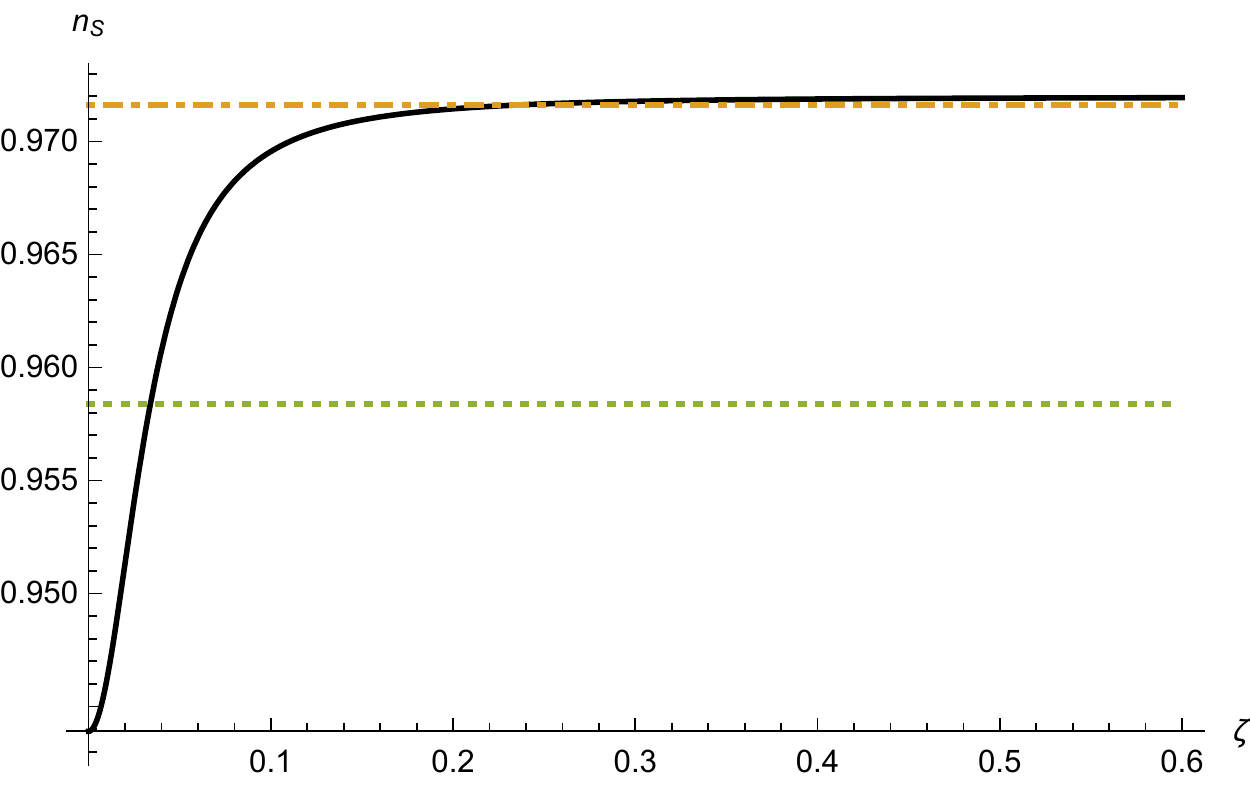}
\includegraphics[width=21pc]{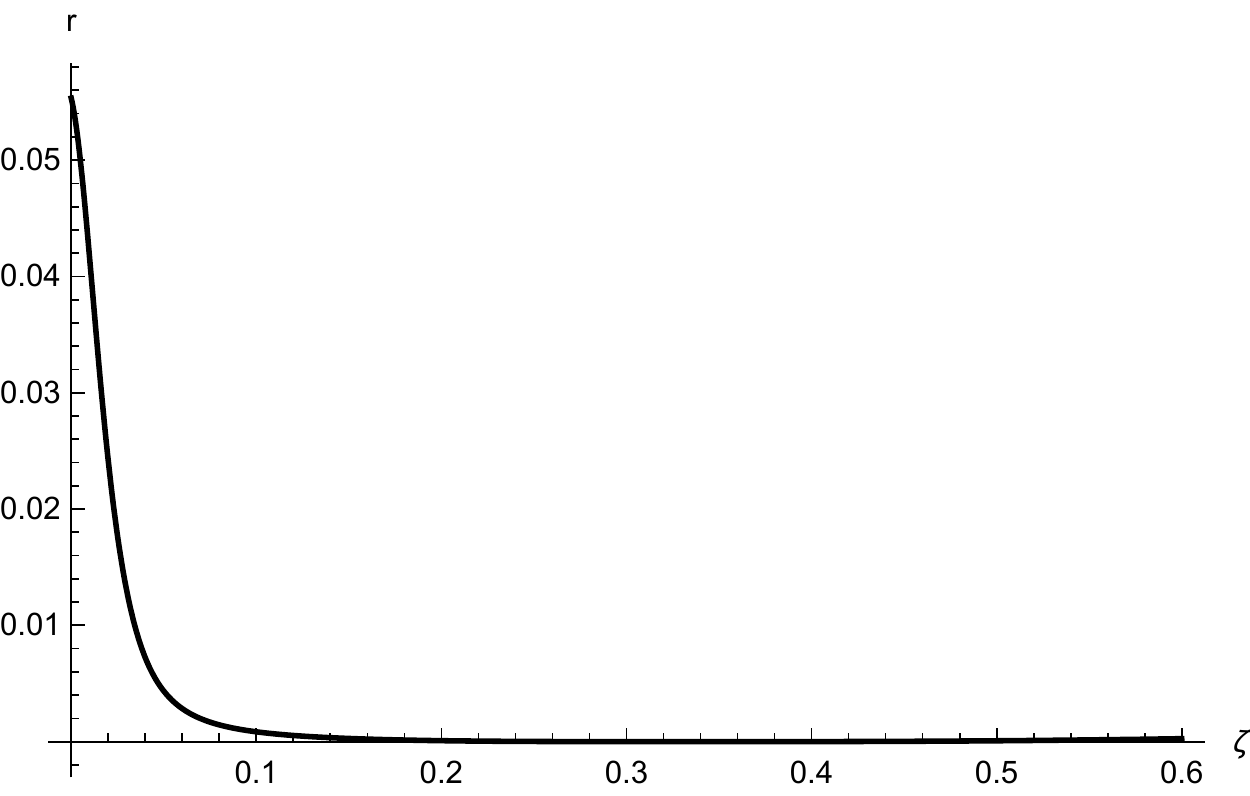}
\caption{Behavior of the  observational indexes $n_s$ and $r$ as functions of the
parameter $\zeta$. In the range $\zeta=[0.034,0.234]$, the observational indexes are compatible with the Planck 2015 \cite{Ade:2015lrj} and BICEP2/Keck-Array data \cite{Array:2015xqh}, which constrain the spectral index and the tensor-to-scalar ratio to satisfy $n_s=[0.95839, 0.97161]$ and  $r<0.07$, respectively.}\label{plot1}
\end{figure}

\begin{figure}[h]
\centering
\includegraphics[width=25pc]{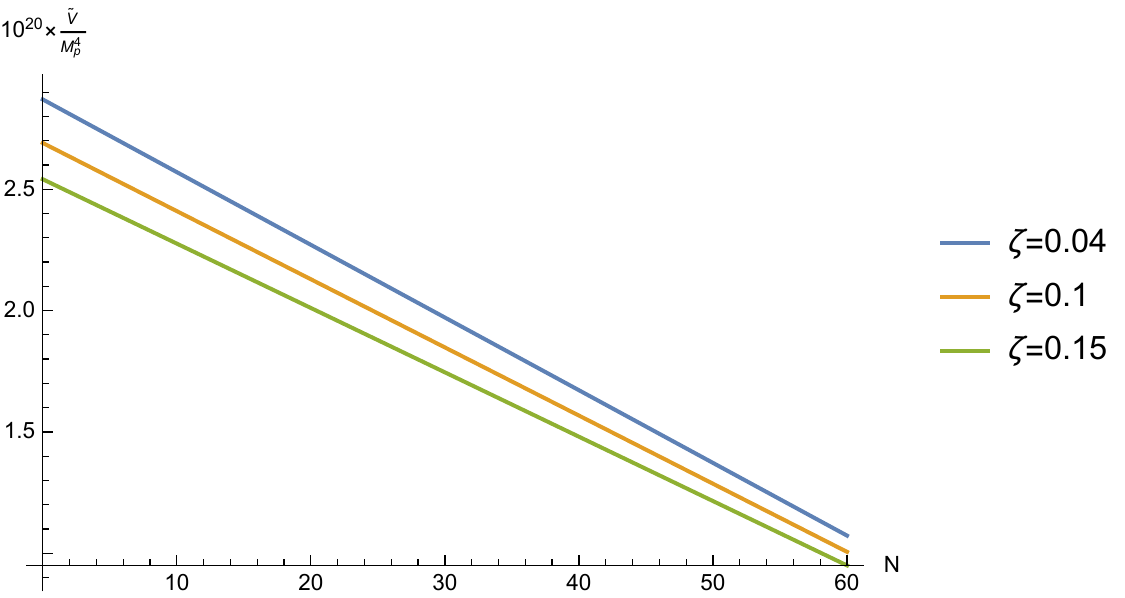}
\caption{Behavior of the potential (\ref{potquasi}) as function of the number of e-folds for the example (\ref{kinconst}) with various values of the parameter $\zeta$ compatible with observations.}\label{plot1v}
\end{figure}

Cosmological perturbation theory based on generalized gravity theories, including string theory correction terms and applications to generalized slow-roll inflations and its consequent power spectra, were studied in Refs. \cite{Noh:2001ia,Hwang:2005hb,Hwang:2002fp,Kaiser:2013sna,sergeipert}. We here adopt the notation and formalism of these works. The observational indexes are quantified in terms of the slow-roll parameters, which in the case of the action (\ref{modelser}) are defined as follows
\[{\varepsilon _1} \equiv \frac{{\dot H}}{{{H^2}}},\,\,\,\,\,\,{\varepsilon _2} = 0,\,\,\,\,\,\,{\varepsilon _3} = 0,\,\,\,\,\,\,\,{\varepsilon _4} \equiv \frac{{\dot E}}{{2HE}},\] 
where the function $E$ is given by
\[E =  - \frac{1}{{{\kappa ^2}}}\lambda (\chi ) + \frac{{3{Q_a}^2}}{{\frac{2}{{{\kappa ^2}}} + {Q_b}}} + {Q_c},\]
and the functions $Q_a$, $Q_b$ and $Q_c$ stand for
\[\begin{gathered}
  {Q_a} = 8{\mu ^2}{h^\prime }(\chi ){H^2} - 4{\mu ^4}{F_1}(\chi )H, \hfill \\
  {Q_b} = 16{\mu ^2}{h^\prime }(\chi )H - 2{\mu ^4}{F_1}(\chi ), \hfill \\
  {Q_c} = 6{\mu ^4}{F_1}(\chi ){H^2}. \hfill \\
\end{gathered} \]
The spectral index of the primordial curvature perturbations, $n_s$, and the tensor-to-scalar ratio, $r$, become respectively \cite{Noh:2001ia,Hwang:2005hb,Hwang:2002fp},
\begin{equation}\label{ns}
{n_s} = 1 + 2\frac{{(2{\varepsilon _1}-{\varepsilon _2}+{\varepsilon _3} - {\varepsilon _4})}}{{1 + {\varepsilon _1}}}
\end{equation}
and
\begin{equation}\label{r}
r = 4\left| {\left( {{\varepsilon _1}-{\varepsilon _3} - \frac{{{\kappa ^2}}}{4}\left( {\frac{1}{{{H^2}}}(2{Q_c} + {Q_d}) - \frac{1}{H}{Q_e} + {Q_f}} \right)} \right)\frac{1}{{1 + \frac{{{\kappa ^2}{Q_b}}}{2}}}{{\left( {\frac{{{c_A}}}{{{c_T}}}} \right)}^3}} \right| ,
\end{equation}
where the functions $Q_d$, $Q_e$ and $Q_f$ read
\[\begin{gathered}
  {Q_d} = 4{\mu ^4}{F_1}(\chi )\dot H, \hfill \\
  {Q_e} = 32{\mu ^2}{h^\prime }(\chi )\dot H - 4{\mu ^2}({\mu ^4}{F_1}'(\chi ) - 2{\mu ^2}{F_1}(\chi )H), \hfill \\
  {Q_f} =  - 16({\mu ^4}{h^{\prime \prime }}(\chi ) - {\mu ^2}{h^\prime }(\chi )H) - 4{\mu ^4}{F_1}(\chi ),\hfill \\
\end{gathered} \]
and the speed of propagation for the scalar and tensor modes, $c_A$ and $c_T$, are given by
\[\begin{gathered}
  c_A^2 = 1 + {\frac{{{Q_d} + \frac{{{Q_a}{Q_e}}}{{\frac{2}{{{\kappa ^2}}} + {Q_b}}} + {Q_f}\left( {\frac{{{Q_a}}}{{\frac{2}{{{\kappa ^2}}} + {Q_b}}}} \right)}}{{ - {\mu ^4}\lambda (\chi ) + \frac{{3{Q_a}^2}}{{\frac{2}{{{\kappa ^2}}} + {Q_b}}} + {Q_c}}}^2}, \hfill \\
  c_T^2 = 1 - \frac{{{Q_f}}}{{\frac{2}{{{\kappa ^2}}} + {Q_b}}}. \hfill \\ 
\end{gathered} \]
In order to calculate $n_s$ and $r$, we consider the inflationary scenario of the quasi de Sitter Eq. (\ref{quasi}). In terms of the Hubble rate, the e-fold number $N$ is equal to
\[N = \int\limits_{{t_k}}^t {Hdt}  = {H_0}(t - {t_k}) - \frac{{{H_i}}}{2}{(t - {t_k})^2}.\]
Solving with respect to $t-t_k$, we get  \cite{reviews1}
\begin{equation}\label{tn}
t - {t_k} = \frac{{{H_0} + \sqrt {{H_0}^2 - 2{H_i}N} }}{{{H_i}}}.
\end{equation}
It is worth noticing that this expression is valid whenever  $N < \frac{{H_0^2}}{{2{H_i}}}$, which gives the maximum number of $e$-folds as $N_{max}= \frac{{H_0^2}}{{2{H_i}}}$.  In the following, just for simplicity, we take
\begin{equation}\label{redefield}
\chi  = {\mu ^2}(t - {t_k}).
\end{equation}
Now  we can calculate the observational indexes by combining Eqs. (\ref{potquasi}), (\ref{lagquasi}), (\ref{ns}), (\ref{r}), (\ref{tn}), and (\ref{redefield}), for arbitrarily chosen functions $F_{1}(\chi)$ and $h(\chi)$. We can consider some explicit examples. As a first, simple one:
\begin{equation}\label{kinconst}
h(\chi ) = 0,\,\,\,\,\,\,\,\,\,{F_1}(\chi ) = {\kappa ^2}\zeta,  
\end{equation}
where $\zeta$ is a dimensionless constant. If we assume that the total number of e-folds is $N = 60$, and also choose ${M_p} = 1/\kappa $, ${H_i} = 0.5 \times {10^{ - 22}}M_p^2$, ${H_0} = {10^{ - 10}}{M_p}$, and $\mu  \sim {M_p}$, we find that the compatibility with the observational data is indeed achieved for values of  the parameter $\zeta$ in the range $[0.034,0.234]$. To illustrate the behavior of the observational indexes, in Fig. \ref{plot1} we plot  $n_s$ and $r$ as functions of the parameter $\zeta$. For this example, in Fig. \ref{plot1v} we also plot the potential (\ref{potquasi}) as function of the number of e-folds for various values of the parameter $\zeta$ compatible with observations.

\begin{figure}[h]
\centering
\includegraphics[width=18pc]{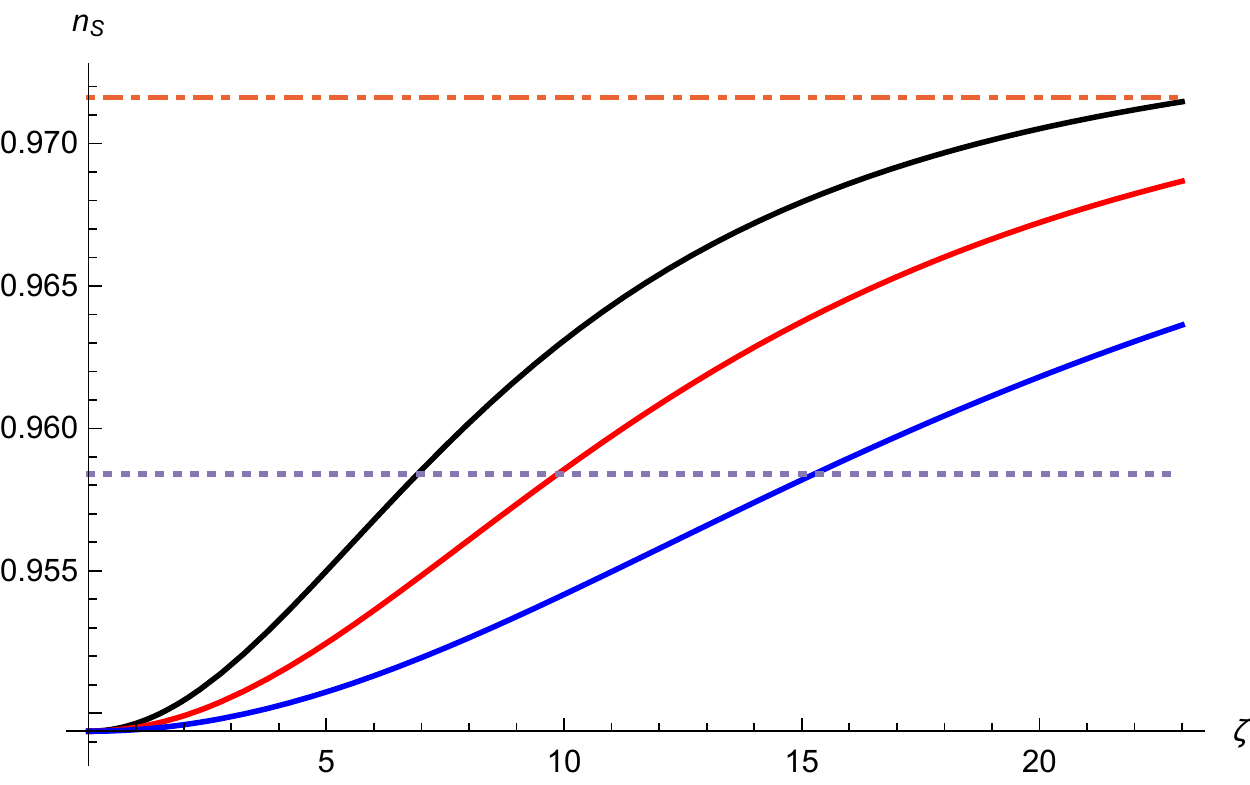}
\includegraphics[width=21pc]{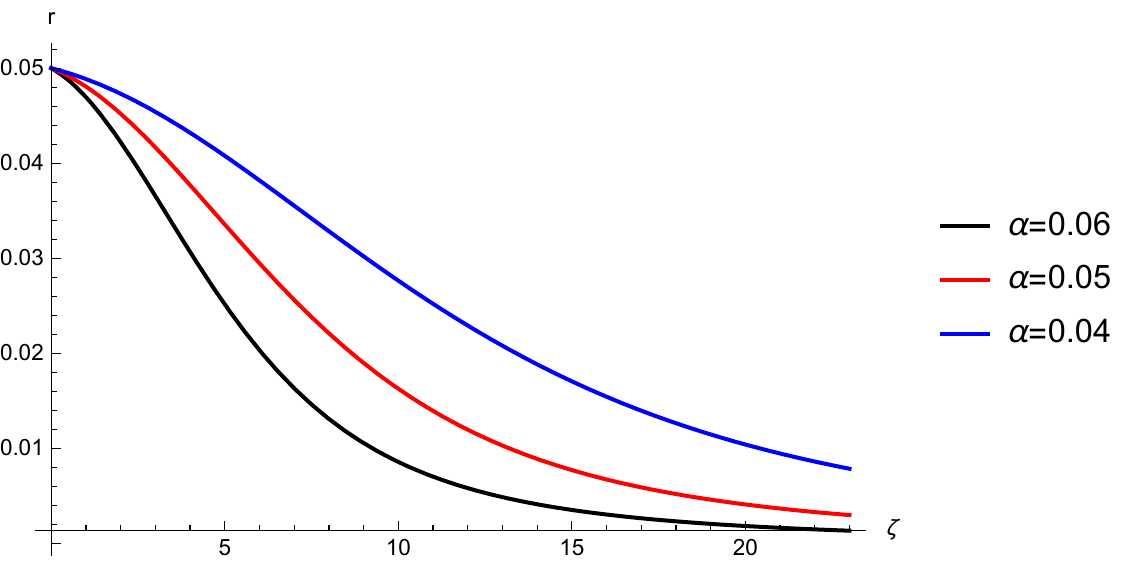}
\caption{Behavior of the  observational indexes $n_s$ and $r$ as functions of the
parameter $\zeta$, for various values of the parameter $\alpha$. The curves cover different behaviors of the observational indexes, all of them being in an acceptable  value range, according to
the data. As is clear from the plot, the simultaneous compatibility of $n_s$ and $r$ with the observational data can be achieved for a wide range of values of the free parameters ($\zeta$, $\alpha$)}\label{plot2}
\end{figure}

\begin{figure}[h]
\centering
\includegraphics[width=25pc]{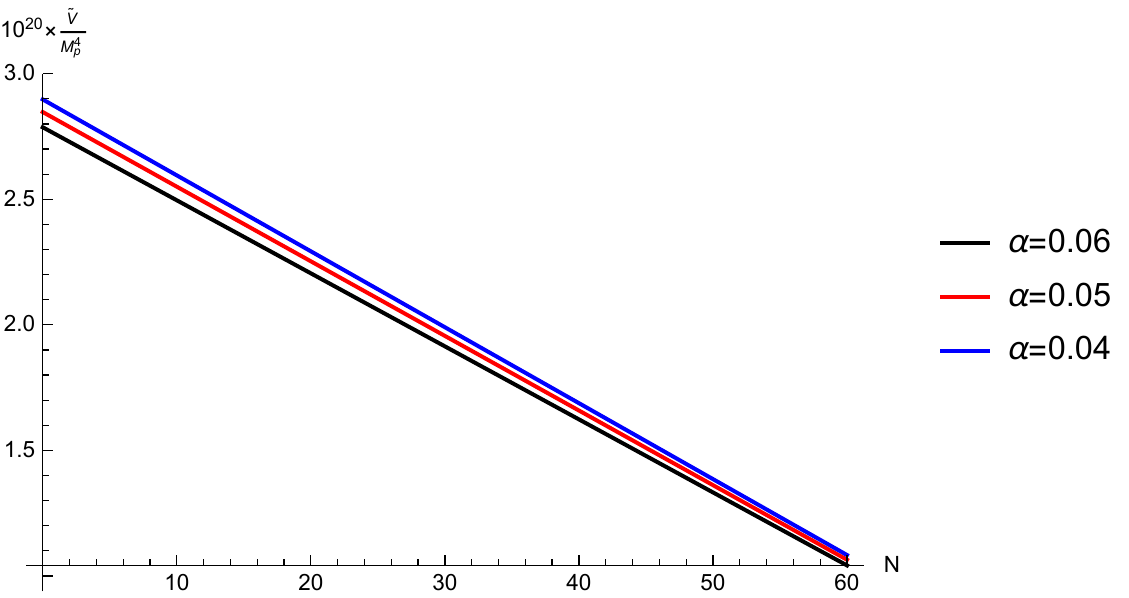}
\caption{Behavior of the potential (\ref{potquasi}) as function of the number of e-folds for the example (\ref{notrivial}) for various values of the parameter $\alpha$  with  $\zeta=20$, which is compatible with observations.}\label{plot2v}
\end{figure}

Another example is:
\begin{equation}\label{notrivial}
h(\chi ) = 0,\,\,\,\,\,\,\,\,\,\,{F_1}(\chi ) = {\kappa ^2}\zeta {\left( {1 - {e^{ -\alpha \tanh \left( {\kappa \chi } \right)}}} \right)^2}
\end{equation}
where $\zeta$ and $\alpha$ are dimensionless constants. After a thorough examination of the parameter space, it is apparent that this case is also compatible with both the Planck \cite{Ade:2015lrj} and BICEP2/Keck-Array data \cite{Array:2015xqh}, for a large range of values of the free parameters. In order to illustrate this fact in a more transparent way, in Fig. \ref{plot2} we present the plot of the spectral index $n_s$ and of the tensor-to-scalar ratio $r$, as functions of the parameter $\zeta$, for various values of the parameter $\alpha$. In this case, the behavior of the potential (\ref{potquasi}) as function of the number of e-folds is plotted in the Fig. \ref{plot2v} for various values of the parameter $\alpha$  with $\zeta=20$, which is compatible with observations.

\section{Conclusions}

In this paper we have extended the theoretical framework of the ghost-free Gauss-Bonnet theories \cite{ghostfree} to other modified gravities, involving a non-minimal kinetic coupling. We have briefly reviewed the essential features of the Lagrange multiplier formalism in this specific model. In our study we have considered an approach that allows, in principle,
to reconstruct the potential and the Lagrange multiplier for arbitrarily given cosmological scenarios. 

We have then considered explicit examples of the reconstruction procedure, namely for power-law, Little-Rip, de Sitter, and quasi de Sitter models, where the resulting potentials depend of arbitrary functions $h(\chi)$ and $F_{1}(\chi)$, which can be chosen in a convenient way such that the phenomenological constraints can be fully satisfied. In particular, we have investigated the role of the non-minimal kinetic coupling in the dynamics of inflation and we have investigated some specific choices for the function $F_{1}(\chi)$. 

After briefly discussing the general formalism of the observational indexes in modified gravity, we have fixed the Hubble rate (quasi de Sitter) and the kinetic coupling. We have thus demonstrated that the resulting inflationary evolution is viable, in a rigorous phenomenological context, by calculating the slow-roll indexes and the corresponding observational indexes, focusing on the spectral index of the primordial curvature perturbations and on the scalar-to-tensor ratio. Moreover, we have proven that viability of the model can be achieved for a wide range of values of the free parameters, and we have illustrated this by displaying various plots of the observational indexes as functions of the free parameters.

Summing up, we have proven here that, in principle, any given early and late-time cosmological solution relevant to current observations (power-law, Little Rip, de Sitter,  quasi de Sitter) can be reconstructed using the model of (\ref{modelser}). We have also shown that the effects of the non-minimal kinetic coupling lead to viable inflationary scenarios, fully compatible with the most recent and accurate observational constraints. Moreover, the choice of the Hubble rate is in principle arbitrary, with the only constraint being that the resulting theory should be an accelerating one, that is, the resulting scale factor must satisfy $\ddot a > 0$. To finish, it is also possible to perform the same analysis for other choices of the Hubble rate during inflation.

\section*{Acknowledgments}

\noindent This work was partially supported by Universidad del Valle under project CI 71074, by MINECO (Spain), FIS2016-76363-P,  the CPAN Consolider Ingenio 2010 Project, and  AGAUR (Catalan Government), project 2017-SGR-247. D.F.J. acknowledges support from COLCIENCIAS, and thanks the Institute of Space Sciences (IEEC-CSIC), UAB and ICREA, Barcelona for kind hospitality.

\end{document}